\documentclass[12pt]{article}
\usepackage{url}
\newcommand{\beq}{\begin{equation}}
\newcommand{\eeq}{\end{equation}}

\begin{document}
\title{
Errors in Low and Lapsley's article "Optimization Flow Control, I: Basic Algorithm
      and Convergence" }
\author{Andrzej Karbowski\\
NASK (Research and Academic Computer Network),\\
ul. W¹wozowa 18,\\
02-796~Warsaw, Poland,\\
E-mail: A.Karbowski@ia.pw.edu.pl}

\date{August 2002 }
\parindent 2em
\maketitle
\thispagestyle{empty}
\begin{abstract}
In the note two errors in Low and Lapsley's article \cite{Low1} are shown.
Because of these errors the proofs of both theorems presented in the article
are incomplete and some assessments are wrong.
\end{abstract}

\section{Error in the proof of the Theorem 1}

In the proof of  Lemma 3 (at the beginning) on the page 871 it is written:
{\em "Given any $p,q \geq 0$, using Taylor theorem and Lemma 2 we have
$\nabla D(q) - \nabla D(p) = \nabla^2 D(w) (q-p) =$ [...] for some $w = t p+(1-t) q \geq 0, t \in [0,1]$."}

The problem is, that the mentioned Taylor theorem, or rather multidimensional
mean value theorem, is not true.
Here is the counterexample:\\
Let us denote
\[
z=\left[ \begin{array}{c}
x \\
y
\end{array} \right]
\]
and take:
\beq
D(z) = x^2 y + y^4
\eeq
and
\[
p = \left[ \begin{array}{c}
0\\
 0
 \end{array} \right],\;\;
q = \left[ \begin{array}{c}
1\\
1
\end{array} \right]
 \]
We will get the following gradient of the function $D(z)$:
\beq
\nabla D(z) =
\left[ \begin{array}{c}
 2xy \\
  x^2 + 4y^3
\end{array} \right]
\eeq
and the following Hessian:
\beq
\nabla^2 D(z) = \left[ \begin{array}{cc}
 2y        &       2x   \\
  2x       &      12 y^2  \\
\end{array} \right]
\eeq
Then, making use of the mentioned "theorem",
denoting $v = 1-t$ we will get:
\beq
\left[ \begin{array}{c}
2 \\
5
\end{array} \right] = \left[ \begin{array}{cc}
2 v     &          2 v        \\
2 v    &    12 v^2
\end{array}
\right] \left[ \begin{array}{c}
1\\
1
\end{array} \right]
\eeq
Hence, at the same time it must be $4v = 2$ and $2v + 12 v^2 = 5$, what is impossible.

\section{Error in the proof of the Theorem 2}

The last passage in the assessment (29) on the page 873 (the proof of Lemma 6) is incorrect.
Low and Lapsley justify it: {\em "where the last inequality holds because the convex
function $\sum_i y_i^2 + z^2 - \sum_i  y_i  z$ attains a unique
minimum over $\{(y_i,z) \,| \;y_i \geq 0, z  \geq 0 \}$ at the origin."}\\
But this is not true.
It is sufficient to take $\dim y = 5$ and $ y = [5,\, 4,\, 3,\, 4,\, 5]$, $ z = 10$.
The considered function takes the value $-19$.\\
It is so, because the Hessian of the considered function has the form
(taking: $x~=~[y_1, y_2, \ldots, z]'$):
\beq
H = \left[ \begin{array}{cccccc}
 2 & 0 &0 &\ldots &0 & -1 \\
 0 & 2 & 0 &\ldots & 0 &-1 \\
 0 & 0 & 2 &\ldots & 0 &-1 \\
 \vdots\\
  0 & 0 & 0 &\ldots & 2 &-1 \\
  -1 & -1 & -1 & \ldots& -1 & 2
 \end{array}\right]
 \eeq
and its characteristic polynomial  (e.g. calculated from the
Schur's formula: for $A~:~n~\times~n;$\, $D~:~m~\times~m,\; B^T,C:m \times n; \;det([A, B; C, D]) =  det(A) \cdot det(D-CA^{-1}B)$\, )  will be:
\beq
det(H - \lambda I) = (2 - \lambda)^{n-1} (\lambda^2 - 4 \lambda + 4 -n)
\eeq
where $n = \dim y$. In this way we will have  eigenvalues: $\lambda_i = 2,\, i=1,\ldots,n-1$
and $\lambda_{n,n+1}~=~2~\pm~\sqrt{n}$. This means, that for $n>4$ there will be one
negative eigenvalue and the function will not  be convex.

\section{Corrections}
The error in the proof of the first theorem was noticed independently by
Edward Fan from UCLA in March 2002. The correction is based on
 the Theorem 9.19 from the
Rudin's book \cite{rudin}.

The correction of the proof of the second theorem  was done by
the author and is on the Web page:\\
\url{http://www.ia.pw.edu.pl/~karbowsk/pub/papers}\\
This paper was submitted to "IEEE/ACM Transactions on Networking"
in August 2002.

According to these suggestions Steven Low corrected his article.
The corrected version is now on the Web page:\\
\url{http://netlab.caltech.edu/pub/papers/ofc1_ToN.pdf}


\begin{thebibliography}{99}

\bibitem{Low1} Low, S. and D.E. Lapsley, Optimization Flow Control, I: Basic Algorithm
      and Convergence, {\em IEEE/ACM Transactions on Networking},
      7(6), pp. 861-874, 1999.
\bibitem{rudin} Rudin, W., {\em Principles of Mathematical Analysis}, McGraw-Hill, Inc.,
\end{thebibliography}
\end{document}